\documentclass[twocolumn,english,aps,prl,nofootinbib]{revtex4}

\usepackage[T1]{fontenc}
\usepackage[latin9]{inputenc}
\usepackage{amsmath}
\usepackage{amssymb}
\usepackage{appendix}
\usepackage{esint}
\usepackage{graphicx}
\usepackage{subfigure}
\usepackage{natbib}
\makeatletter
\@ifundefined{textcolor}{}
{%
 \definecolor{BLACK}{gray}{0}
 \definecolor{WHITE}{gray}{1}
 \definecolor{RED}{rgb}{1,0,0}
 \definecolor{GREEN}{rgb}{0,1,0}
 \definecolor{BLUE}{rgb}{0,0,1}
 \definecolor{CYAN}{cmyk}{1,0,0,0}
 \definecolor{MAGENTA}{cmyk}{0,1,0,0}
 \definecolor{YELLOW}{cmyk}{0,0,1,0}
 }

\usepackage{nicefrac}\usepackage{dsfont}
\usepackage[normalem]{ulem}\@ifundefined{definecolor}
 {\usepackage{color}}{}

\newenvironment{additionalnotes}{\begin{description}}{\end{description}}

\makeatother

\usepackage{babel}
\begin{document}

\title{
Strain-controlled criticality governs the nonlinear mechanics of fibre networks}
\author{A.\ Sharma\textsuperscript{1,2}, A.J.\ Licup\textsuperscript{1}, R.\ Rens\textsuperscript{1}, M.\ Sheinman\textsuperscript{1}, K.A.\ Jansen\textsuperscript{3}, G.H.\ Koenderink\textsuperscript{3}, F.C.\ MacKintosh\textsuperscript{1}}
\address{\textsuperscript{1}Department of Physics and Astronomy, VU University, Amsterdam, The Netherlands\\
\textsuperscript{2} III. Physikalisches Institut, Georg-August-Universit{\"a}t, 37077 G{\"o}ttingen, Germany\\
\textsuperscript{3}FOM Institute AMOLF, Science Park 104, 1098 XG Amsterdam, The Netherlands}

\date{\today}
\begin{abstract}

\end{abstract}

\maketitle

\textbf{
Disordered fibrous networks are ubiquitous in nature as major structural components of living cells and tissues.
The mechanical stability of networks generally depends on the degree of connectivity: only when the average number of connections between nodes exceeds the \emph{isostatic} threshold are networks stable~\cite{maxwell1864}. Upon increasing the connectivity through this point, such networks undergo a mechanical phase transition from a floppy to a rigid phase. However, even sub-isostatic networks become rigid when subjected to sufficiently large deformations.
To study this strain-controlled transition, we perform a combination of computational modeling of fibre networks and experiments on networks of type I collagen fibers, which are crucial for the integrity of biological tissues. We show theoretically that the development of rigidity is characterized by a strain-controlled continuous phase transition with signatures of criticality. Our experiments demonstrate mechanical properties consistent with our model, including the predicted critical exponents. We show that the nonlinear mechanics of collagen networks can be quantitatively captured by the predictions of scaling theory for the strain-controlled critical behavior over a wide range of network concentrations and strains up to failure of the material.}

As shown by Maxwell, networks with only central-force interactions exhibit a transition from a floppy to rigid phase at the \emph{isostatic} point, where the local coordination number, or connectivity $\langle z \rangle$ equals the threshold value of $\langle z \rangle = 2d$ in $d$ dimensions~\cite{maxwell1864}. At this point, the number of degrees of freedom is just balanced by the number of constraints, and the system is \emph{marginally stable} to small deformations. The jamming transition~\cite{cates1998jamming,liu1998nonlinear,majmudar2007jamming,van2010jamming,saarloos2010jamming} in granular materials and rigidity percolation~\cite{thorpe1983continuous,feng1984percolation,jacobs1995generic,latva2001rigidity} in disordered spring networks are examples of such a transition. An important feature of these systems is the order of the transition. Jamming exhibits signatures of both first- and second-order transitions, with discontinuous behaviour of the bulk modulus and continuous variation of the shear modulus~\cite{olsson2007critical,head2009critical,van2010jamming}. For networks of springs or fibres, the transition from floppy to rigid is a continuous phase transition, in both bulk and shear moduli~\cite{thorpe1983continuous,wyart2008elasticity,ellenbroek2009non,van2010jamming,chase2011,sheinman2012nonlinear}.

Interestingly, the structural networks in biology almost always have connectivities below the central-force isostatic point. Networks such as the extracellular matrices of collagen that make up tissue are strictly sub-isostatic with respect to central forces: their typical connectivity is between 3 (local branching) and 4 (binary crosslinking), placing them below both 2D and 3D isostatic thresholds~\cite{lindstrom2010bio, licup2015stress}. Such sub-isostatic networks can, nevertheless, become rigid as a result of other mechanical constraints, such as fibre bending~\cite{Head2003PRL,wilhelm2003elasticity,chase2011}, or when subjected to external strain~\cite{alexander1998amorphous}. The threshold strain, at which the transition occurs, depends on the nature of applied deformation (shear or tensile) and on the average connectivity of the network, in particular, and other properties of its structure~\cite{sheinman2012nonlinear}. However, the order of this strain-induced transition remains unclear.

\begin{figure}
\includegraphics[width = \columnwidth]{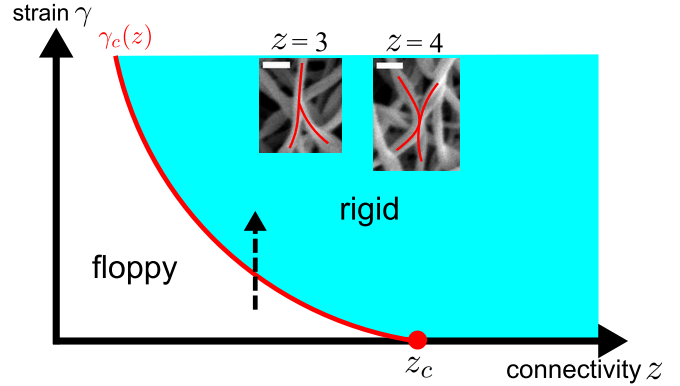}
\caption{At zero strain, $\gamma$, networks undergo a continuous transition from floppy to rigid at the isostatic threshold $z=z_c$. This connectivity threshold shifts to lower values for networks subject to shear strain $\gamma$. This threshold defines a line $\gamma_c(z)$ of continuous transitions. We study here strain-induced transitions indicated by the vertical dashed line for $z$ well below $z_c$. The insets show SEM (Scanning Electron Microscope) images of reconstituted collagen networks indicating points of 3-fold and 4-fold connectivities. The scale bar corresponds to 200 nm.}
\label{PhDiag}
\end{figure}

\begin{figure*}[t]
\centering     
\includegraphics[width = \textwidth]{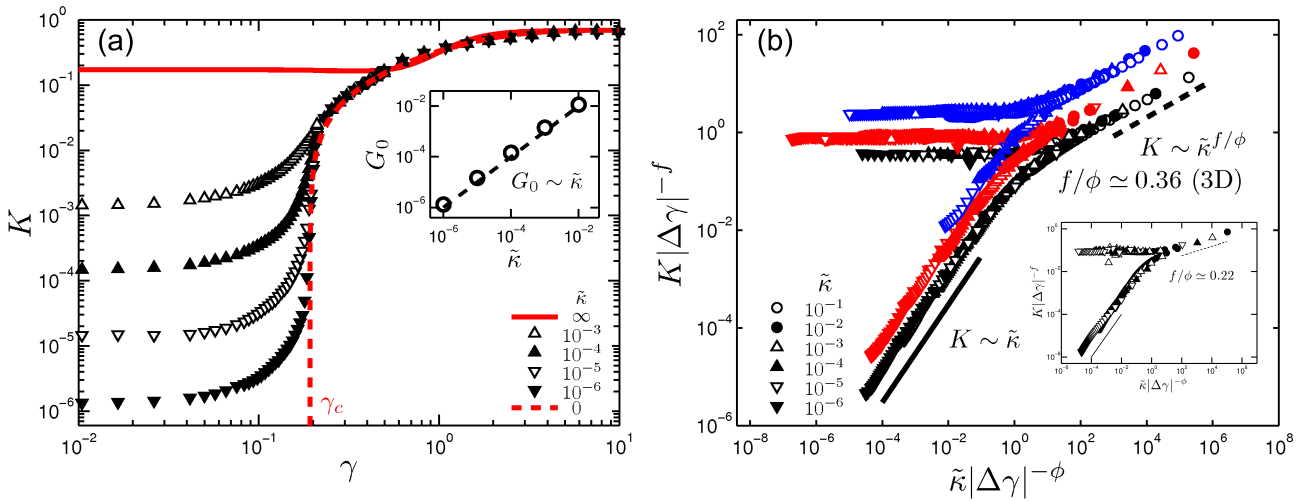}
\caption{(a) Stiffness $K$ in units of $\mu/l$ vs.\ strain $\gamma$ for 2D phantom triangular lattices with $\langle z\rangle \simeq 3.2$. 
The red dashed curve, starting from $\gamma = \gamma_c$, is a sketch of the stiffness of a sub-isostatic network with $\tilde{\kappa} = 0$ while the red solid curve is the affine limit with $\tilde{\kappa} = \infty$. The inset shows the $\tilde\kappa$-dependence of the linear modulus $G_0$ with the dashed line indicating a unit slope. (b)  Collapse of stiffening curves for different $\tilde{\kappa}$. Black curve: the data set from (a) collapsed according to Eq.~\eqref{collapse}, $f = 0.75$ and $\phi = 2.1$. Blue curve: 2D Mikado network with 
$\langle z \rangle\simeq3.6$, $f = 0.84$ and $\phi = 2.2$. Red curve: 3D phantom FCC lattice with $\langle z \rangle \simeq 3.2$, $f = 0.8$ and $\phi = 2.2$. In 3D, $K$ is measured in units of $\mu /l^2$. The solid line has unit slope and the dashed line has slope $f/\phi = 0.38$. The inset shows the data collapse according to Eq.~\eqref{collapse} for a disordered honeycomb lattice with $\langle z \rangle \simeq 2.4$ with critical exponents $f = 0.48$ and $\phi = 2.2$.
}
\label{stiffeningcurves}
\end{figure*}

Here, we study the transition from floppy to rigid states of disordered sub-isostatic networks under simple shear. We show that these networks exhibit a line of second order transitions (see Fig.\ \ref{PhDiag}) at a strain threshold $\gamma_c(z)$, for connectivities $z$ well below the isostatic threshold. Moreover, we demonstrate critical behaviour along this line, specifically in the scaling properties of the mechanics, as well finite-size effects that reflect the underlying divergent correlation length. To test the relevance of these predictions for real materials, we perform experiments on reconstituted networks of collagen, the most prevalent protein in mammals and the mechanical basis of most tissues~\cite{fratzl2008collagen}. Although collagen has been widely studied for many years, the mechanical properties of collagen matrices remain poorly understood. We find that collagen networks show evidence of critical behaviour in their mechanical response to strain. Strikingly, the measured shear modulus of these networks is in quantitative agreement with the critical behaviour of our model, including the predicted non mean-field critical exponents.

We study computational models (See Methods) of fibre networks, based on both 2D and 3D lattice-based structures~\cite{broedersz2011molecular,broedersz2012filament} and Mikado networks in 2D~\cite{wilhelm2003elasticity,conti2009cross}. All networks are, by construction, sub-isostatic and floppy in the absence of bending interactions~\cite{chase2011}. The filaments have a stretching modulus, $\mu$, and bending modulus, $\kappa$. These parameters define a dimensionless rigidity $\tilde\kappa = \kappa/\mu l^2$, where $l$ is the lattice spacing (mesh size) in lattice-based (Mikado) networks. The networks are subjected to simple shear strain $\gamma$ and allowed to relax by minimization of the total elastic energy per unit volume, $\mathcal{H}$, which is calculated using a discrete form of the extensible wormlike chain Hamiltonian~\cite{head2003distinct}. The stress and stiffness, in units of $\mu/l^{d-1}$, are obtained from $\mathcal{H}$ by $\sigma=\tfrac{d\cal{H}}{d\gamma}$ and ${K}=\tfrac{d^2 \cal{H}}{d\gamma^2}$, respectively.

\begin{figure*}[t!]
\includegraphics[width = \textwidth]{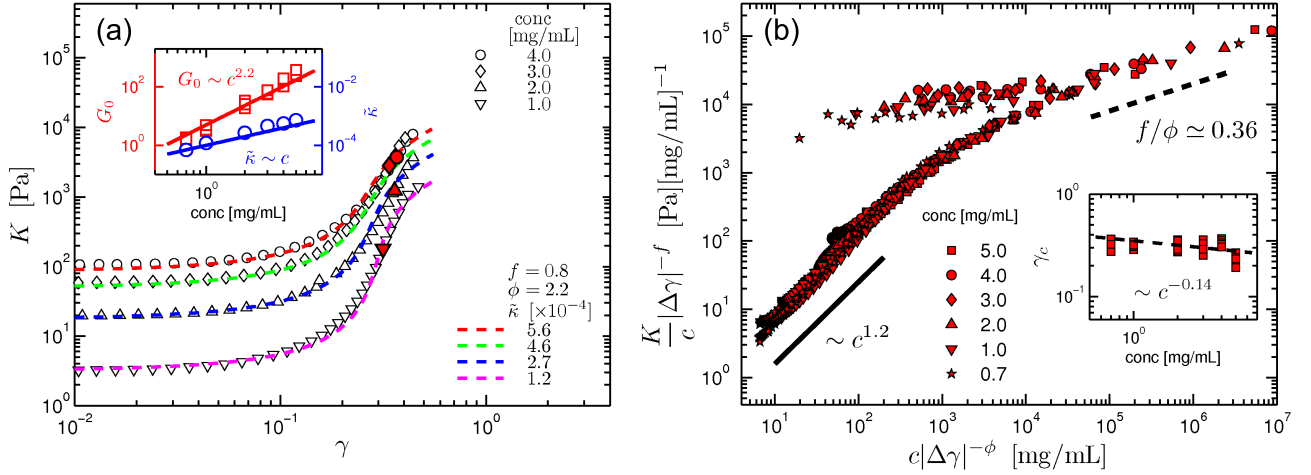}
\caption{(a) Nonlinear stiffness vs. strain measured for collagen networks prepared at different concentrations (see legend).
The inflection point ($\gamma_c$) is in each case marked in red enlarged symbols. The dashed lines are the prediction of Eq.~\eqref{crossover}. The fit values of $\tilde{\kappa}$ based on Eq.~\eqref{crossover} are shown in the inset. The linear scaling of $\tilde{\kappa}$ with the concentration $c$ is consistent with the predictions of our model (Supplementary Information). Also shown in the inset is the linear modulus $G_0$ scaling with concentration as $c^{2.2}$. (b) Collapse of experimental stiffening curves for different concentrations of collagen (see legend). Solid line has slope $1.2$ and the dashed line $f/\phi = 0.36$. Inset shows the weak dependence of critical strain on collagen concentration. The dashed line represents the model prediction of $\gamma_c \sim c^{-0.14}$.}
\label{widomcollapse}
\end{figure*}

In Fig.~\ref{stiffeningcurves}, we show the network stiffness ${K}$ vs.\ strain $\gamma$ of a triangular network with $\langle z \rangle \simeq 3.2$ for different values of $\tilde{\kappa}$.
As sketched in Fig.\ \ref{PhDiag}, these networks are characterized by a continuous transition at a strain threshold $\gamma_c$, which is indicated in Fig.~\ref{stiffeningcurves}a by the vertical dashed red line, above which the stiffness $K$ increases continuously from zero for $\tilde\kappa=0$. This curve is approached for systems with finite but decreasing $\tilde\kappa$, as can be seen by the lower sets of black data points.
This second-order phase behaviour is qualitatively analogous to the onset of ferromagnetism on lowering the temperature below the Curie temperature, where the addition of a magnetic field results in finite magnetization in the paramagnetic phase. More precisely, an additional energy~\cite{wyart2008elasticity} such as the elastic bending stiffness of fibres~\cite{chase2011} with a finite coupling constant $\kappa$ can stabilize otherwise floppy networks. As we show below, the stabilizing effect of $\kappa$ can be used to reveal the critical behaviour for sub-isostatic systems with $z<z_c$ at strains $\gamma\simeq\gamma_c(z)$.

In the absence of the stabilizing effect of bending, i.e., for $\tilde{\kappa}=0$, the continuous nature of the transition in $\gamma$ is apparent (Supplementary Fig.~S2) in the critical scaling of the network stiffness ${K}\sim|\Delta\gamma|^f$ in the regime where $\Delta\gamma=\gamma-\gamma_c>0$. For $\gamma<\gamma_c$, the effect of stabilization by bending leads to ${K}\sim\tilde\kappa$. These regimes can be summarized by the scaling form
\begin{equation}
{K} \propto \left|\Delta \gamma\right|^f\mathcal{G_{\pm}}\left({\tilde\kappa}/{|\Delta \gamma|^{\phi}}\right),
\label{collapse}
\end{equation}
where $\mathcal{G_{\pm}}$ is a scaling function with the positive and negative branches corresponding to $\Delta \gamma >0$ and $\Delta \gamma <0$, respectively. This scaling is analogous to that for the conductivity of random resistor networks and fibre networks as a function of connectivity ~\cite{straley1976critical,chase2011}. In Fig.~\ref{stiffeningcurves}b, we test this scaling relation by plotting ${K}|\Delta\gamma|^{-f}$ vs. $\tilde\kappa|\Delta\gamma|^{-\phi}$, according to Eq.~\eqref{collapse}. For $x\ll 1$, $\mathcal{G_+}(x)$ is approximately constant and $\mathcal{G_-}(x)\propto x$. Since $K$ must be finite at $\Delta \gamma = 0$, we also expect ${K}\sim \kappa^{f/\phi}\mu^{1-f/\phi}$ as one observes from the critical branch in Fig.~\ref{stiffeningcurves}b, consistent with Eq.~\eqref{collapse}. To show the generality of this result, we also show in Fig.~\ref{stiffeningcurves}b the corresponding data obtained from Mikado networks, as well as FCC-based 3D lattices. Strikingly, the data collapse with similar exponents $f=0.8\pm0.05$ and $\phi=2.1\pm0.1$. The average connectivity for the three different networks is chosen to be in the range $\simeq 3.2$-$3.6$, comparable to typical biopolymer networks such as collagen~\cite{lindstrom2010bio}.

To test these predictions for networks of collagen type I, we measure the stiffness vs strain over a range of concentrations using shear rheometry (See Methods) as shown in Fig.~\ref{widomcollapse}a. We also measured the average coordination number of these networks, $z=3.3 \pm 0.1$, consistent both with previous studies~\cite{lindstrom2010bio} and the values used above in our model.
The parameter $\tilde{\kappa}$ in our model is naturally related to the protein concentration $c$ (See Fig.~\ref{widomcollapse}a and Supplementary Information) such that by rescaling our experimental $K$ by the concentration we can compare with the simulations. In both experiment and simulation, we obtained the critical strain $\gamma_c$ as the inflection point of the $\log K$ vs.\ $\log \gamma$ curves (Supplementary Fig.~S1a). This strain also coincides with the strain at which the non-affine fluctuations in the network diverge (Supplementary Fig.~S1b). Based on these considerations, our model predicts that the experimental stiffness $K$ should be governed by the scaling relation
\begin{equation}
{K/c}\propto \left|\Delta \gamma\right|^f\mathcal{G_{\pm}}\left({c}/{|\Delta \gamma|^{\phi}}\right).
\label{exptcollapse}
\end{equation}
In Fig.~\ref{widomcollapse}b, we test this prediction by plotting $K|\Delta \gamma|^{-f}/c$ vs.\ $c|\Delta \gamma|^{-\phi}$. We find an excellent collapse with our predictions for 3D networks shown in Fig.~\ref{stiffeningcurves}b, assuming exponents $f = 0.8$ and $\phi = 2.1$.

In the limit of $\gamma \rightarrow 0$, using $\mathcal{G_{-}}(x) \sim x$ (Eq.~\eqref{collapse}), we obtain a scaling relation for the linear modulus and the concentration: $K/c \sim c \gamma_c^{(f - \phi)}$.
Since $\gamma_c$ is expected to depend on the average connectivity, which may vary with concentration, we expect $\gamma_c$ to show a possible concentration dependence. In fact, as seen in the inset of Fig.~\ref{widomcollapse}b, the experimental $\gamma_c$ does show a weak $\gamma_c \sim c^{-0.14}$ dependence on the concentration. Moreover, this dependence is
consistent with the observed difference between the experimental and theoretical $\mathcal{G_{-}}$ branches of Fig.~\ref{stiffeningcurves}b and ~\ref{widomcollapse}b: $K/c \sim c^{1.2}$ is consistent with $K/c \sim c \gamma_c^{(f - \phi)}$ and the exponents  $f$ and $\phi$ in Fig.~\ref{widomcollapse}b, since $\gamma_c \sim c^{0.2/(f-\phi)} \sim c^{-0.14}$.

To quantitatively capture the mechanics of fibrous networks over the entire strain-range, one needs the scaling function $\mathcal{G}_{\pm}(x)$. As for ferromagnetic systems, one can obtain an approximate scaling function by numerical inversion of the equation of state~\cite{ArrottPRL1967} (Supplementary Information), which in our case is given by
\begin{equation}
\frac{\tilde{\kappa}}{|\Delta \gamma|^{\phi}} \sim \frac{{K}}{|\Delta \gamma|^f}\left(\pm 1+\frac{{K}^{1/f}}{|\Delta \gamma|}\right)^{(\phi - f)}.
\label{crossover}
\end{equation}
Here, $\pm$ corresponds to the two branches of the scaling function. In Fig.~\ref{widomcollapse}a, we show the predicted ${K}$ vs.\ $\gamma$ according to Eq.~\eqref{crossover}, with $\tilde{\kappa}$ as the only fitting parameter (Supplementary Information). 

\begin{figure}[t]
\includegraphics[width = \columnwidth]{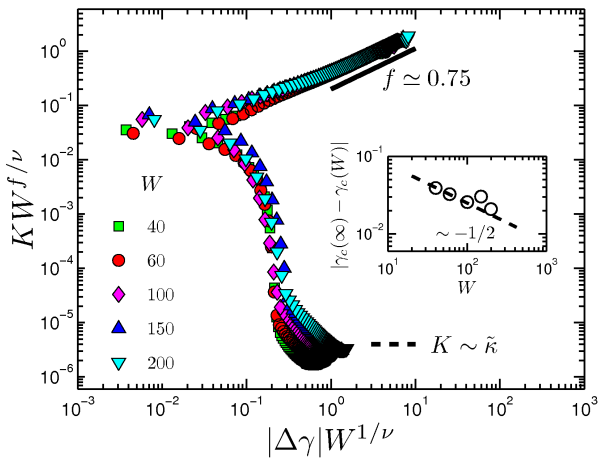}
\caption{Demonstration of the continuous transition of elasticity in a sub-isostatic network. Collapse of stiffening curves according to Eq.~\eqref{kcollapse} for different system sizes $W$. Inset shows the scaling of $|\gamma_c(\infty) - \gamma_c(W)|$ with system size. The dashed line has a slope of $-1/2$.}
\label{FSS}
\end{figure}

We further test the critical behaviour by performing finite-size scaling, which is sensitive to the divergence of the correlation length. The modulus should follow the scaling relation (Supplementary Information)
\begin{equation}
{K} \propto W^{-{f}/{\nu}}\mathcal{F_{\pm}}\left(|\Delta \gamma| W^{{1}/{\nu}}\right),
\label{kcollapse}
\end{equation}
where $W$ is the system size and $\mathcal{F_{\pm}}$ is a scaling function with the positive and negative branches corresponding to $\Delta \gamma >0$ and $\Delta \gamma <0$, respectively. In Fig.~\ref{kcollapse}, we plot ${K} W^{f/\nu}$ vs. $|\Delta \gamma|W^{1/\nu}$, showing consistency with Eq.\ (\ref{kcollapse}). These data were obtained for 2D lattice-based networks with $\tilde{\kappa} = 10^{-7}$. We obtain a good collapse of the data for $f = 0.75 \pm 0.05$ and $\nu = 2.0 \pm 0.1$. The lower branch, $\mathcal{F_{-}}$ does not vanish as $W \rightarrow \infty$, due to the small, finite value of $\tilde\kappa$. We confirm, however, that $\mathcal{F_{-}}$ for $W\gtrsim1$ decreases toward zero as $\tilde{\kappa}\rightarrow0$ (Supplementary Fig.~S3).

In Table \ref{exptable} we summarize the various values of $f$ and $f/\phi$ for different network structures in either 2D or 3D.  Although the theoretical results above were chosen to correspond to connectivity $\langle z \rangle$ close to the experimental values for collagen networks, we also studied two very different networks---a disordered honeycomb lattice in 2D with $\langle z \rangle$ close to 2 (See Fig.~\ref{stiffeningcurves}b inset) and a disordered FCC lattice with $\langle z \rangle$ close to (Supplementary Fig. S6).
Importantly, the near-isostatic case of the FCC lattice at $\langle z \rangle \simeq5$ exhibits a value $f/\phi\simeq1/2$, consistent with the $K\sim\kappa^{0.5}$ scaling reported in Ref.~\cite{chase2011} for an isostatic network.
We note, however, that the individual exponents $f$ and $\phi$ here are only defined for sub-isostatic networks, and thus are not expected to coincide with studies of isostatic systems~\cite{wyart2008elasticity,chase2011}.


\begin{table}
\centering
\caption {Critical exponents obtained by simulations of networks varying in connectivity, architecture, and dimensionality.}
  \begin{tabular}{ l | c | c | r  }
  \hline
    $\langle z\rangle$ & Network & $f$ & $f/\phi$ \\ \hline
    2.4 & 2d Honeycomb & 0.5 & 0.22 \\ 
    3.2 & 2d phantom & 0.75 & 0.36 \\ 
    3.2 & 3d phantom& 0.8 & 0.36 \\ 
    3.6 & 2d Mikado& 0.84 & 0.4 \\ 
    5.0 & 3d FCC& 1.45 & 0.5 \\ \hline
  \end{tabular}
  \label{exptable}
 \end{table}

In contrast to most prior work on critical phenomena for jamming~\cite{van2010jamming}, rigidity percolation~\cite{thorpe1983continuous,feng1984percolation} and near-isostatic fibre networks~\cite{wyart2008elasticity,chase2011}, as well as predictions of  topological boundary modes in isostatic lattices~\cite{kane2014topological}, our focus here has been on networks well below the isostatic point. This situation is particularly relevant to biology, which abounds with structural networks networks of biopolymers with connectivity between 3 and 4, well below isostaticity in 3D. One challenge in understanding such systems has been their nonlinear mechanical response. Recently, a Landau type theory for the non-linear elasticity of biopolymer gels was proposed using an order parameter describing induced nematic order of fibres in the gel~\cite{feng2014alignment}. Our findings show that the nonlinear mechanics of networks of stiff fibres such as collagen can now be understood quantitatively in terms of critical phenomena associated with the isostatic point. Importantly, as we show, there is a line of critical points that extends over a wide range of network connectivities, covering the physiologically relevant range of $3<z<4$ in 3D. Moreover, although the linear modulus of collagen networks may be finite in this range due to the stabilizing influence of bending, the nonlinear response can be quantitatively captured by the scaling functions in Eqs.\ (\ref{collapse},\ref{exptcollapse}).

\begin{additionalnotes}
\item[Supplementary Information] accompanies this paper at www.nature.com/naturephysics.
\item[Acknowledgments] We thank M. Vahabi for useful discussions. This work is
part of the research programme of the Foundation for Fundamental Research on Matter (FOM),
which is financially supported by the Netherlands Organisation for Scientific Research (NWO).
This work is further supported by NanoNextNL, a micro and nanotechnology programme of the
Dutch Government and 130 partners.
\item[Author Contributions] AS and AJL contributed equally to the work. AS, AJL, RR, MS, and FCM conceived and developed the model and simulations. AS, AJL, RR and MS performed the simulations. KAJ and GHK designed the experiments. KAJ performed the experiments. All authors contributed to the writing of the paper.
\item[Additional Information] The authors declare no competing financial interests. Reprints and permission information is available online at http://npg.nature.com/reprintsandpermissions/

Correspondence should be addressed to GHK or FCM.
\end{additionalnotes}

\section*{Methods Summary}
\subsection*{Network generation}
We model lattice-based networks in 2D and 3D, as well as off-lattice (Mikado) networks in 2D. In our lattice-based networks, fibres are arranged on a triangular lattice (2D) or a face-centered cubic (FCC) lattice (3D) of linear dimension $W$. In 2D, we randomly select two of the three fibres at each vertex on which we form a binary cross-link, i.e., enforcing local 4-fold connectivity of the network in which the third fibre does not interact with the other two. In 3D, where there are 6 fibres crossing at a point, we randomly connect three separate pairs of fibres at each vertex with binary cross-links to enforce local 4-fold connectivity. In both 2D and 3D, the average connectivity is further reduced below 4 by random dilution of bonds with a probability ($1-p_{\mathrm{bond}}$). In order to generate 3D networks with connectivity exceeding 4, we simply perform random dilution of bonds on a full FCC lattice, with an initial connectivity of 12, till the desired connectivity is reached. The stretching modulus $\mu$ and the lattice constant $l_0$ is set to 1 in 2D and 3D. Mikado networks are generated by random deposition of filaments in a 2D box of size $W$. A freely hinged cross-link is inserted at every point of intersection. The deposition continues until the desired average connectivity is obtained.

\subsection*{Collagen Rheology}
For our experiments, rat-tail collagen type I (BD Biosciences, Breda) was polymerized at 37 degrees Celsius, in a in a physiological buffer solution composed of DMEM1x solution (diluted from 10x, Sigma) containing 50 mM HEPES, 1.5 mg/ml sodium bicarbonate, 1\% FBS (Gibco) and 0.1\% antibiotics (pen/strep, Gibco) at pH 7.3. Rheology was performed on a stress-controlled rheometer (Physica MCR 501; Anton Paar, Graz, Austria) with a 40 mm cone-plate geometry having an 1 degree cone angle. A solvent trap was added to maintain a humid atmosphere. After 6 hours polymerization, stress-stiffening curves were obtained using a differential protocol.



\def\bibsection{\subsection*{\refname}}

\renewcommand{\figurename}{\textbf{Supplementary Figure}} 
\def\fnum@figure{\figurename\nobreakspace\textbf{\thefigure}}
\setcounter{figure}{0}\renewcommand{\thefigure}{S\arabic{figure}}

\renewcommand{\theequation}{S\arabic{equation}}

\pagenumbering{gobble}

\section*{Supplementary Information}
\subsection*{Critical strain as the inflection point of strain-stiffening curves}
\begin{figure*}[t!]
\centering
\includegraphics[width=\textwidth]{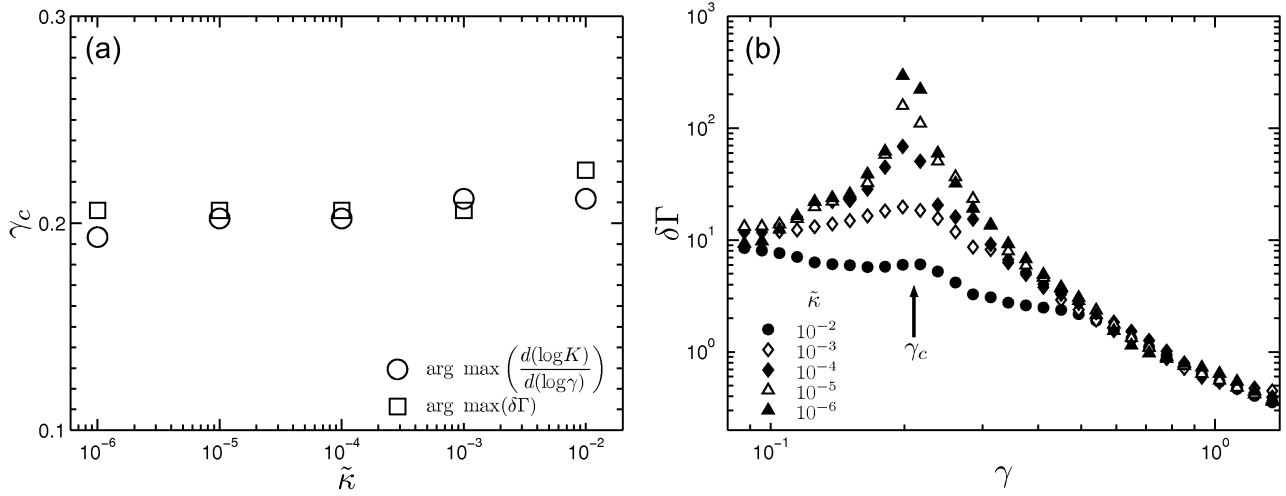}
\caption{(a) Critical strain $\gamma_c$, obtained from simulations on a 2D phantom triangular network with $\langle z \rangle \simeq 3.2$, as the inflection point of the $\log K$ vs.\ $\log \gamma$ curves (circles). Squares show the strain at which the non-affine fluctuations show a maximum. It is clear that the maximum in the non-affine fluctuations occurs at $\gamma \simeq \gamma_c$. (b) Differential non-affinity $\delta \Gamma(\gamma)$ obtained from the same simulations in (a) as a function of the applied strain. $\delta \Gamma(\gamma)$ peaks at $\gamma \simeq \gamma_c$. As expected, the height of the peak increases with decreasing $\tilde{\kappa}$ since the displacement field of these networks becomes highly non-affine as $\gamma\rightarrow\gamma_c$.}
\label{gcdna}
\end{figure*}

We define the critical strain $\gamma_c$ as the inflection point of the $\log K$ vs.\ $\log \gamma$ curves, analogous to the determination of the critical point in a finite size system~\cite{stauffer1994introduction}. Using this definition, we can unambiguously extract the critical strain in a consistent manner for both experiments and simulations. In Fig.~\ref{gcdna}a, we show the critical strain, obtained from simulations on a 2D phantom triangular network with $\langle z \rangle \simeq 3.2$, for different values of the fibre bending rigidity. As can be seen, $\gamma_c$ shows a weak dependence on the $\tilde{\kappa}$. In the limit of $\tilde{\kappa}\rightarrow 0$, $\gamma_c$ approaches a constant value. Moreover, $\gamma_c$ also marks the strain at which the network exhibits the largest non-affine fluctuations. Given the displacement field $\mathrm{\mathbf{u}}$ and the \emph{affine} displacement field $\mathrm{\mathbf{u}^{A}}$ of the network, the non-affine fluctuations can be quantified as~\cite{sheinman2012nonlinear}
\begin{equation}
\delta \Gamma(\gamma) = \frac{\langle \|\delta \mathrm{\mathbf{u}^{NA}}\|^2 \rangle}{l^2 d\gamma^2},
\label{dna}
\end{equation}
where $\delta \Gamma(\gamma)$ is referred as differential non-affinity, $\delta\mathrm{\mathbf{u}^{NA} = \mathbf{u} - \mathbf{u}^{A}}$ is the differential non-affine displacement of a crosslink to an imposed strain $d\gamma$, $l$ is the typical network mesh size and the angular brackets represent a network average. In fact, in the limit of $\tilde{\kappa}\rightarrow 0$, the fluctuations are expected to diverge. In Fig.~\ref{gcdna}b, we show $\delta \Gamma(\gamma)$ in the neighbourhood of the critical strain. As can be seen, the fluctuations grow with decreasing $\tilde{\kappa}$, consistent with the idea that the bending rigidity can be considered as an auxiliary field.

\begin{figure*}[t!]
\centering
\includegraphics[width=\textwidth]{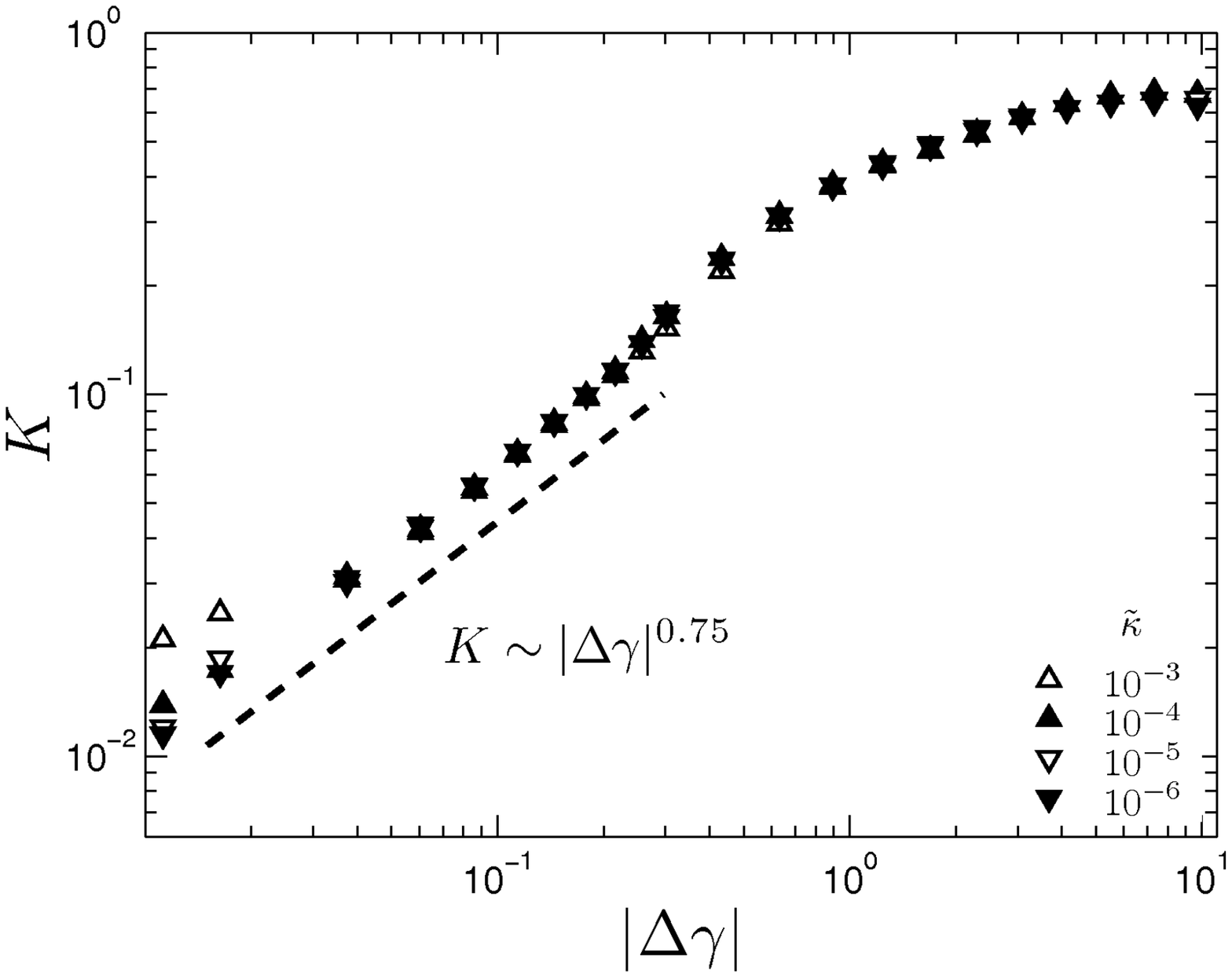}
\caption{Stiffening curves for $\gamma>\gamma_c$ in a 2D network (phantom triangular, $\langle z \rangle \simeq 3.2$) shows that in the limit $\tilde{\kappa} \rightarrow 0$, $K\sim |\Delta\gamma|^f$ where for this network $f\simeq 0.75$.}
\label{abovegammac}
\end{figure*}

\section*{Concentration scales linearly with the reduced bending rigidity $\tilde{\kappa}$}
The parameter $\tilde{\kappa}$ in our model is naturally related to the protein concentration $c$ as follows. For an elastic rod of radius $r$ and Young's modulus $E$, $\mu = \pi r^2 E$ and $\kappa = \pi r^4 E/4$, implying that $\tilde{\kappa} \propto (r/l)^2\propto \phi$, since the volume fraction $\phi=\pi r^2\rho$, where $\rho\propto1/l^2$ is the total fibre length per volume~\cite{head2003distinct,wilhelm2003elasticity,conti2009cross}. Hence, the protein concentration $c$ (or $\rho$) in experiments can be simply related to the reduced bending rigidity $\tilde{\kappa}$ as $\rho \sim \tilde{\kappa}$. The theoretical elastic energy involves a summation over all fibres in the network and is a function of the strain $\gamma$ and the reduced bending rigidity $\tilde{\kappa}$. Moreover, since the modulus $K$ involves the energy per unit volume, $K$ is naturally proportional to $\rho$. The modulus can therefore be expressed as
\begin{equation}
K = \mu \rho \mathcal{K}\left(\gamma, \tilde{\kappa}\right).
\end{equation}
In the linear regime, $\gamma \rightarrow 0$, we obtain from the above the well known relation $K \sim \mu \rho \tilde{\kappa} \sim \rho^2$.

\subsection*{Scaling analysis}
\subsubsection*{Finite size scaling}
\begin{figure*}[t!]
\centering
\includegraphics[width=\textwidth]{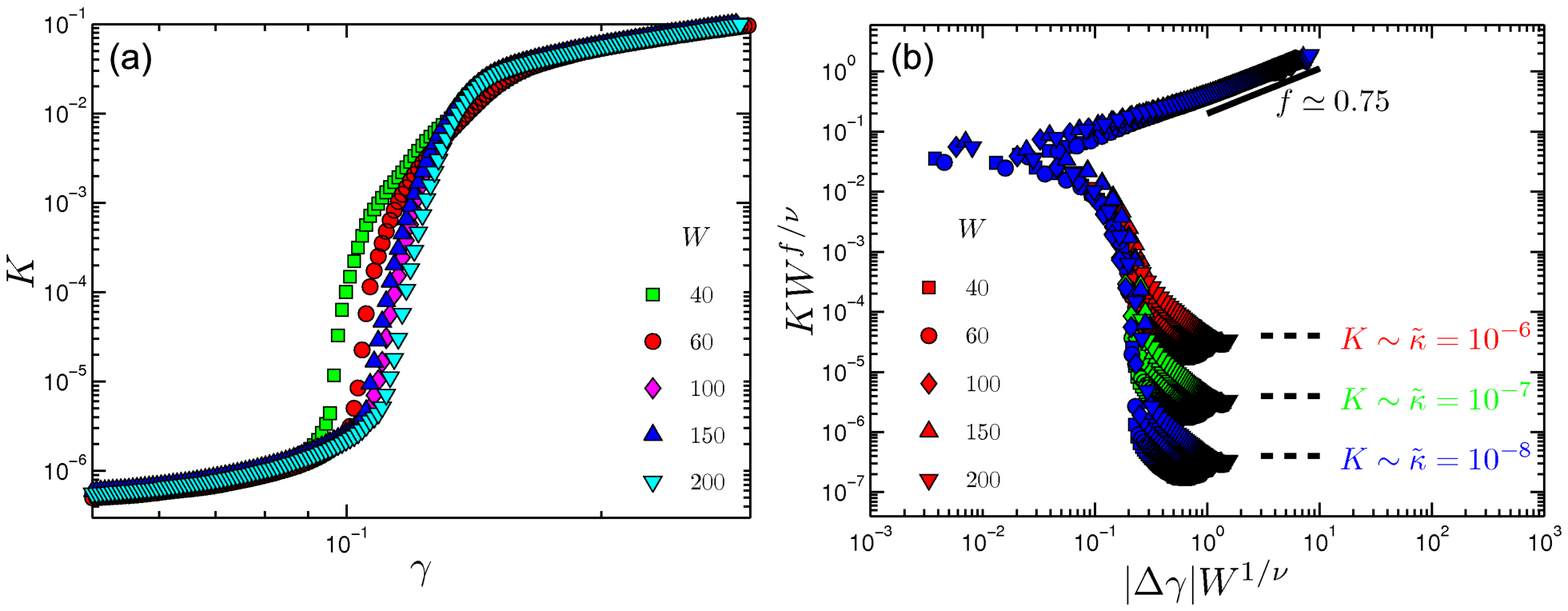}
\caption{(a) Stiffening curves for $\tilde{\kappa} = 10^{-7}$ for different system sizes. (b) Collapse of the stiffening curves according to Eq.~\eqref{kcollapse} for different system sizes $W$. Stiffening curves are obtained for three different bending rigidities (see legend). The lower branch converges to a value which scales with the bending rigidity $\tilde{\kappa}$. In the limit of $\tilde{\kappa} \rightarrow 0$, the lower branch will extend all the way down to zero.}
\label{fsscontinuity}
\end{figure*}

The critical behavior in our model can be tested by performing finite-size scaling, which is sensitive to the divergence of the correlation length. In our system, the order parameter $K$ scales as $|\Delta \gamma|^f$ (Fig.~\ref{abovegammac}), as system size $W\rightarrow \infty$ and $\Delta \gamma \rightarrow 0$. This scaling should be evident when the correlation length, which scales as $|\Delta \gamma|^{-\nu}$, is much smaller than the system size, where $\nu$ is the correlation length exponent. At the critical point, when the correlation length diverges, the modulus should scale with system size as $K \sim W^{-f/\nu}$, such that $K \rightarrow 0$ as $W \rightarrow \infty$. This can be summarized in the following scaling relation.

\begin{equation}
K = W^{-{f}/{\nu}}\mathcal{F_{\pm}}\left(|\Delta \gamma| W^{{1}/{\nu}}\right).
\label{kcollapse}
\end{equation}

In order to perform finite-size scaling, we chose a finite but small bending rigidity. We considered system sizes from $W=40$ to a maximum of $W=200$ in 2D. We chose a finite $\tilde{\kappa}$ to avoid the numerical problems associated with a rope-like network ($\tilde{\kappa} = 0$). In a rope-like network, due to finite size effects, the energy density shows a jump at the critical strain. Since one needs to take derivatives of energy density, it is numerically problematic to unambiguously extract the modulus at the critical strain. In Fig.~\ref{fsscontinuity}a we show the stiffening curves for $\tilde{\kappa} = 10^{-7}$ for different system sizes. In Fig.~\ref{fsscontinuity}b we present the collapse (Eq.~\eqref{kcollapse}) of stiffening curves obtained for two other bending rigidities, $\tilde{\kappa} = 10^{-6}$ and $10^{-8}$. As can be seen, for the three values of $\tilde{\kappa}$, the data collapse is indistinguishable except for the lower branch. The lower branch converges to a value which scales with $\tilde{\kappa}$. It follows that the lower branch extends continuously to zero as $\tilde{\kappa} \rightarrow 0$, consistent with the fact that a central-force sub-isostatic network is unstable for $\gamma \leq \gamma_c$.

\subsubsection*{Equation of state}
In ferromagnetism, the magnetization, $m$, in presence of an applied field $h$ and reduced temperature $t = (T-T_c)/T_c$ can be captured in the following scaling relation~\cite{ArrottPRL1967}:
\begin{equation}
\frac{h}{|t|^{\Delta}} \sim \frac{m}{|t|^{\beta}}\left(\pm 1 + \frac{m^{1/\beta}}{|t|}\right)^{(\Delta-\beta)}.
\label{magnetization}
\end{equation}
Here, the $(\pm)$ branch corresponds to $t\gtrless 0 $ and $\Delta$ and $\beta$ are critical exponents. The analogous quantities for a fibre network are the following:
\begin{eqnarray}
\Delta &\leftrightarrow& \phi \\
\beta &\leftrightarrow& f \\
\frac{h}{|t|^{\Delta}} &\leftrightarrow& \frac{\tilde{\kappa}}{|\Delta \gamma|^{\phi}}\\
\frac{m}{|t|^{\beta}} &\leftrightarrow& \ \frac{K}{|\Delta \gamma|^f}
\end{eqnarray}

Based on the above analogy, we obtain the following scaling law.
\begin{equation}
\frac{\tilde{\kappa}}{|\Delta \gamma|^{\phi}} \sim \frac{K}{|\Delta \gamma|^f}\left(\pm 1+\frac{K^{1/f}}{|\Delta \gamma|}\right)^{(\phi - f)}.
\label{crossover}
\end{equation}
It can be seen that for $\Delta \gamma = 0$, the above scaling relation correctly reproduces $K \sim \tilde{\kappa}^{f/\phi}$ at the critical point.

In the main text, we use  Eq.~\eqref{crossover} to obtain fit to the experimental $K$ vs. $\gamma$ data in the following way. We first focus on the linear regime. In the linear regime, we know from simulations that the modulus (in units of $\rho \mu$) scales linearly with $\tilde{\kappa}$ which itself scales as $\tilde{\kappa} \sim \rho$ giving rise to the well known $c^2$ (or $\rho^2$) dependence of the linear modulus where $c$ is the protein concentration. It follows that in order to compare experimental $K$ with that obtained from simulations we should first rescale the experimental $K$  by $c^{1.2}$ so that the rescaled modulus scales as $K/c^{1.2} \sim c \sim \tilde{\kappa}$. As the next step, we obtain the individual critical strains, $\gamma_c$, for each of the concentrations as the inflection point of the $\log K$ vs. $\log \gamma$ curve. We then consider the experimental data (rescaled by $c^{1.2}$) for the lowermost concentration along with its $\gamma_c$ and perform fitting to it using Eq.~\eqref{crossover} with $\tilde{\kappa}$ as the only free parameter. Once we obtain $\tilde{\kappa}$ for a given protein concentration, we obtain the $\tilde{\kappa}$ for other concentrations by using $\tilde{\kappa} \sim c$. However, in the fit shown in the main text, we have treated $\tilde{\kappa}$ as a free parameter. We show in the main text that the fit values of $\tilde{\kappa}$ are consistent with the expected $\tilde{\kappa} \sim c$ scaling.

\subsection*{Average connectivity of collagen networks}
\begin{figure*}
\centering
\includegraphics[width=\textwidth]{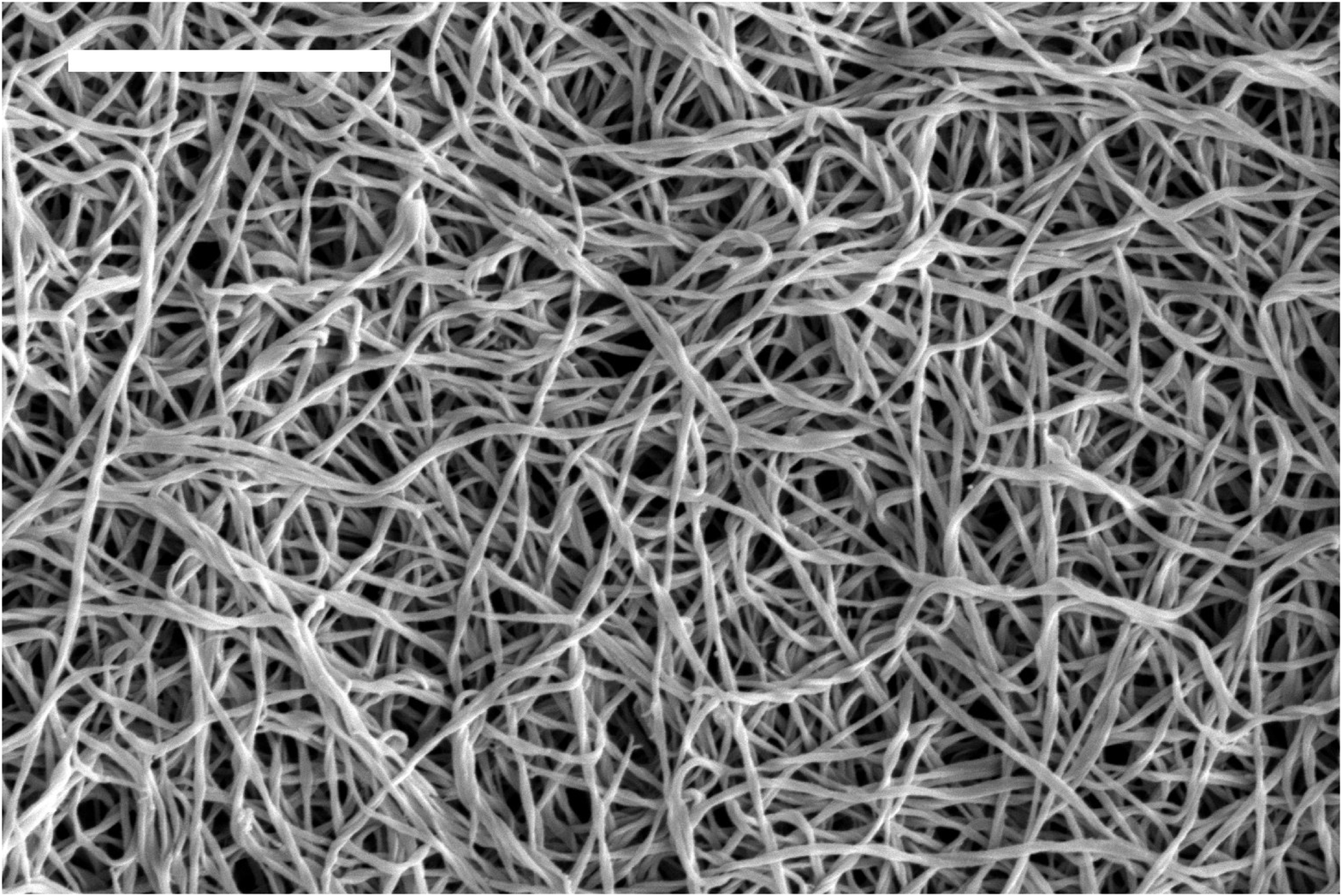}
\caption{SEM image of collagen network prepared at a concentration of 4 mg/ml. The branched structure of the collagen fibres is visible. The white scale bar represents 2 $\mu$m.}
\label{SEM}
\end{figure*}

In Fig.~\ref{SEM}, we show a typical Scanning Electron Microscope (SEM) image of a collagen network prepared at a temperature $T=37^{\circ}$ and concentration of 4 mg/ml. For SEM, collagen gels (50-100 $\mu$l) were polymerized overnight inside 5 ml eppendorf tubes in humid conditions. After polymerization, samples were washed three times with sodium cacodylate buffer (50 mM cacodylate, 150 mM NaCl, pH 7.4) for 30-60 min each, at their polymerization temperature. Samples were fixed with 2.5\% glutaraldehyde in the same buffer for at least 2 hours. Next, samples were washed three times with sodium cacodylate buffer (room temperature) and dehydrated with increasing amount of ethanol. After complete dehydration (100 \% ethanol), 50 \% \emph{hexamethyldisilazane} (HMDS) in ethanol was added (under the hood) and afterwards replaced after 30 min by 100\% HMDS. The HMDS was left to evaporate overnight. The samples were transported to a stub with carbon tape and sputter coated using a K575X sputter coater (Quorum Technologies, Gouda, The Netherlands). A layer of 15.4 nm of Au/Pd was sputtered using a current of 80 mA. The SEM samples were visualized using a Scanning Transmission Electron Microscope (STEM) setup (Verios 460, FEI Company, Eindhoven, the Netherlands) using 50 pA, 5 kV and 4 mm working distance, in immersion mode. Two different samples, with number of branches $>$ 100, concentration of 4mg/ml and temperature $T=37^{\circ}$, were analyzed. The average connectivity of a collagen network with concentration 4mg/ml is measured to be $3.3\pm0.1$.

\begin{figure*}[t!]
\centering
\includegraphics[width=\textwidth]{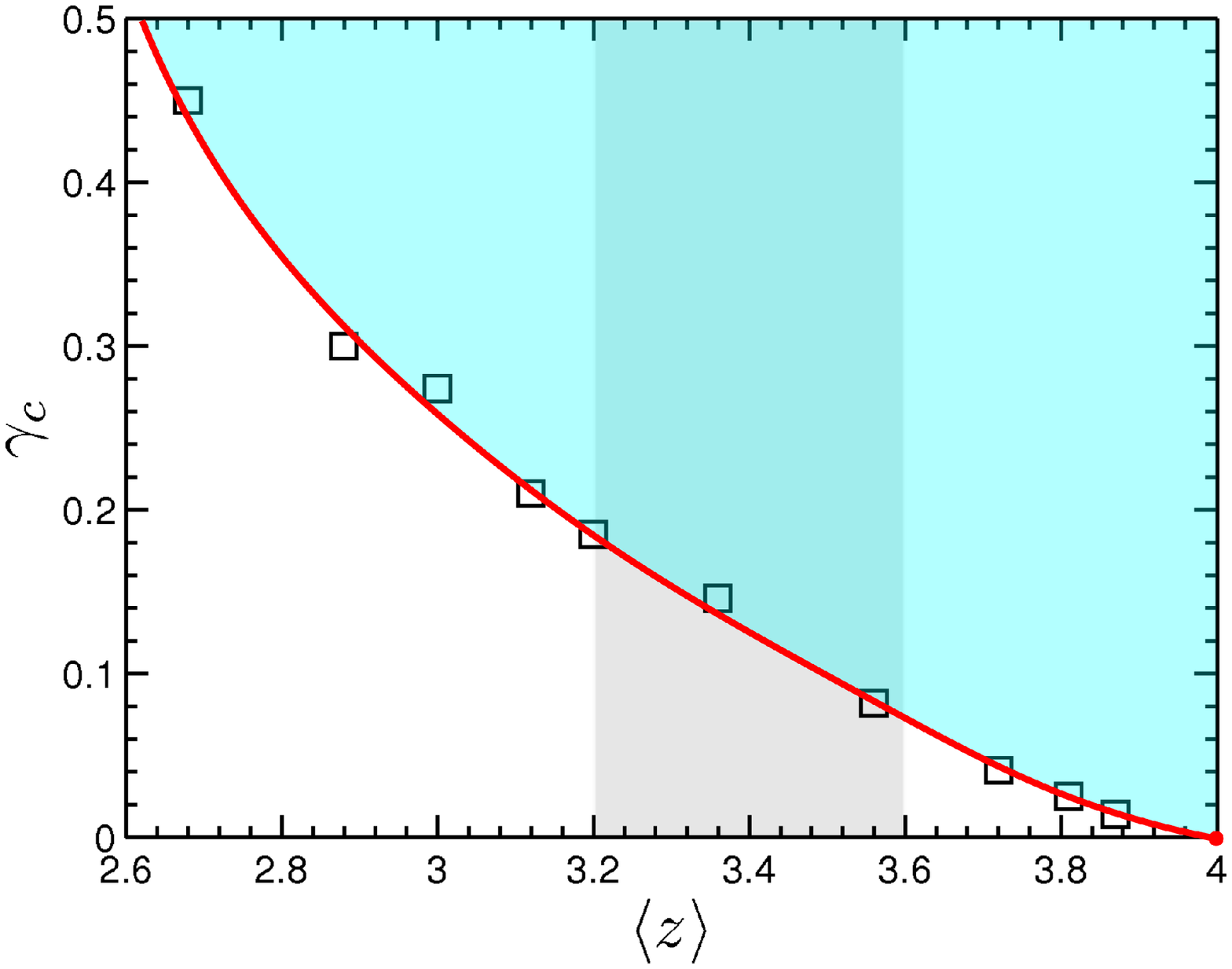}
\caption{Critical strain as a function of the average connectivity in a 2D network (phantom triangular). In the shaded rectangular region, $\langle z \rangle$ changes from 3.2 to 3.6, a range relevant for collagen.}
\label{gcvsz}
\end{figure*}

\subsection*{Critical strain as a function of connectivity}
In Fig.~\ref{gcvsz}, we show the critical strain as a function of average connectivity in a 2D phantom triangular network. In the main text, a schematic version of such a figure is shown as Fig.~1. As can be seen in Fig.~\ref{gcvsz}, $\gamma_c$ approaches zero as the connectivity approaches the isostatic threshold of $\langle z \rangle = 4$ in 2D. In Fig.~\ref{gcvsz}, we mark as shaded rectangle the connectivity-range relevant for collagen networks. In the main text, the overall decrease in the critical strain in the experiments (Fig.~3b inset in the main text) with increasing concentration of collagen is about 30\%. Increasing concentration of collagen is expected to result in an increase in the average connectivity of the network. Consistent with this, one can see in Fig.~\ref{gcvsz} that the critical strain decreases with increasing connectivity. In fact, $\gamma_c$ decreases significantly from $0.2$ to $0.1$, as the connectivity increases from $3.2$ to $3.6$.



\section*{Near isostatic 3D FCC network}
\begin{figure*}[t!]
\centering
\includegraphics[width=\textwidth]{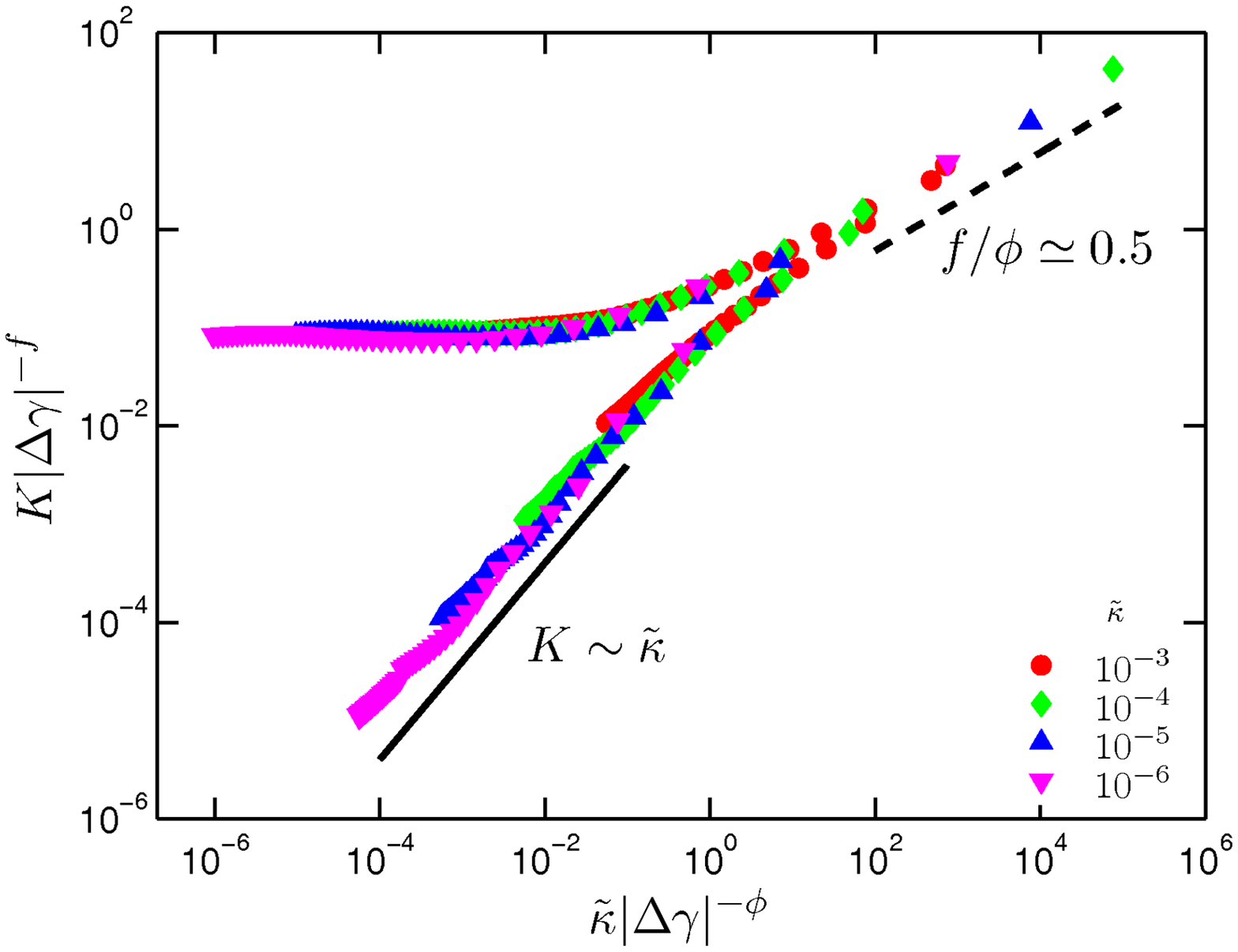}
\caption{Collapse of data from a 3D FCC lattice network with $\langle z \rangle \simeq 5$. The critical exponents are $f = 1.45$ and $\phi = 2.9$. The ratio $f/\phi\simeq1/2$ is consistent with the $K\sim\kappa^{0.5}$ scaling reported in Ref.~\cite{chase2011} for an isostatic network in 3D.}
\label{nearisostatic}
\end{figure*}

Although the theoretical results shown in the main text correspond to connectivity $\langle z \rangle$ close to the experimental values for collagen networks, we also studied a disordered FCC lattice with $\langle z \rangle$ close to 5. The predicted scaling is shown in Fig.~\ref{nearisostatic}. Such a network is close to the isostatic threshold of $\langle z \rangle = 6$ in 3D. The near-isostatic case of the FCC lattice at $\langle z \rangle \simeq 5$ exhibits a value $f/\phi\simeq1/2$, consistent with the $K\sim\kappa^{0.5}$ scaling reported in Ref.~\cite{chase2011} for an isostatic network.
The individual exponents are $f = 1.45$ and $\phi = 2.9$. These critical exponents are only defined for sub-isostatic networks, and thus are not expected to coincide with studies of isostatic systems~\cite{wyart2008elasticity,chase2011}.

\end{document}